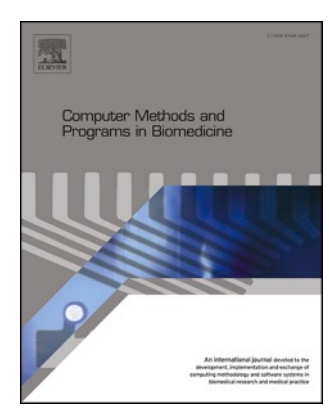

# Impact of friction force and retrieval speed on in silico mechanical thrombectomies: A sensitivity analysis

Mahesh S. Nagargoje [a,*], Virginia Fregona [a], Giulia Luraghi [a], Francesco Migliavacca [a,b], Demitria A Poulos [c], Bryan C Good [c], Jose Felix Rodriguez Matas [a]

[a] *Computational Biomechanics Laboratory, Laboratory of Biological Structure Mechanics (LaBS), Department of Chemistry, Materials and Chemical Engineering "Giulio Natta, Politecnico di Milan, 20133 Milan, Italy*
[b] *Fondazione IRCCS Ca' Granda Ospedale Maggiore Policlinico, Milan, Italy*
[c] *Mechanical, Aerospace, and Biomedical Engineering, University of Tennessee, Knoxville, Knoxville, TN, USA*

A B S T R A C T

*Background:* Mechanical Thrombectomy (MT) is a widely accepted first-line treatment for Acute Ischemic Stroke (AIS) and it has been studied using in vitro and in silico models. Thrombectomy outcomes have been performed for patient-specific cases using in silico models. However, until now, in vivo friction coefficients for stent-vessel, stent-clot, and clot-vessel interactions are unknown, but in vitro experiments have been attempted with significant standard deviations. These interactions and friction coefficients have been considered an important aspect of thrombectomy success.
*Objectives:* In the current study, we explored the influence of variation in friction forces for stent-vessel, stent-clot, and clot-vessel interactions using virtual mechanical thrombectomy (VMT). We have performed three simulations for each interaction and varied friction coefficients around the standard deviation observed in the past in vitro studies.
*Results:* (i) clot-vessel friction: higher friction leads to clot fragmentation and VMT failure. (ii) stent-clot friction: it is susceptible to VMT outcomes, with lower values showing the slippage of the clot while higher values lead to fragmentation. (iii) stent-vessel friction: higher friction shows compression of the stent in curved vessels and dislodgment of clot from stent retriever (SR) due to its compression, which leads to VMT failure. (iv) retrieval speed (RS): higher RS (>30 mm/s) leads to significant stent compression and unrealistic behavior of the SR.
*Conclusions:* Analysis of results proposes the necessity for calculating accurate friction factor values and their implementation into in silico models, due to their sensitivity towards thrombectomy outcomes. Such in silico models mimic in vivo thrombectomy more closely and can be used in mechanical thrombectomy planning, management, and decision-making.

## 1. Introduction

Over one million inhabitants of Europe suffer from a stroke each year, with Acute Ischemic Stroke (AIS) accounting for 80 % of total cases [1]. In 2017, 1.12 million incidents of stroke were recorded in the European Union, (0.46 million resulting in death) and the number of people living with stroke is expected to increase by 27 % by 2047 [2]. Treatment of AIS has been revolutionized by the introduction of mechanical thrombectomy (MT) in clinical practice in 2012 [3]. MT proposes an effective way to remove lodged embolus in the cerebral artery using a stent retriever (SR) or aspiration catheter [4]. The "MR CLEAN" trial confirmed the safety and effectiveness of MT by compiling data from 500 patients [5]. Other randomized control trials proved MT is an effective way of treating AIS [6,7]. Although MT is an effective and successful treatment option for AIS, there remains a sizable proportion of patients who are unsuccessful in reperfusion of the blood flow in the cerebral circulation [4]. The failure rate of thrombectomy has motivated researchers to understand MT mechanics better using in silico and in vitro models.

MT procedures have been studied using in silico and in vitro approaches, with their respective pros and cons [8–14]. In vitro models are simple to set up, but difficult to control clot and vessel properties. In contrast, in silico models are difficult to set up, but easier to vary clot and vessel properties and other variables. In silico models also provide the flexibility of parametric analysis when limited knowledge is available in the literature [14]. For the first time, Virtual Mechanical Thrombectomy (VMT) simulations in a realistic cerebral circulation geometry have been implemented by Luraghi et al. and validated through the comparison with in vitro tests [15,16]. Later, the same methodology was used to replicate a VMT in a patient-specific case [17].

* Corresponding author.
*E-mail address:* maheshshahadeo.nagargoje@polimi.it (M.S. Nagargoje).

Irrespective of the above advances, the friction coefficients used in the past in silico models may vary due to the absence of in vivo data. For the first time, Gunning et al. proposed in vitro friction coefficients for clot-vessel static friction in MT, but dynamic friction coefficients were missing [18]. A recent article by Elkhayyat et al. analyzed static and dynamic friction coefficients for the interactions between artificial surfaces commonly used for vitro models, bovine arterial tissue, SRs, and red blood cell (RBC)-rich and fibrin-rich clots [19]. In the above studies, it was observed that friction coefficients for clot-vessel were higher with fibrin-rich clots in comparison to RBC-rich clots. The friction was also higher in the case of arterial tissue in comparison with artificial model surfaces. Further, the dynamic friction between SRs and vessel walls was observed to be significantly and needs to be included in the in silico models. The standard deviations of the friction coefficients in the above studies, however, were significantly high and a better understanding of the sensitivity of these parameters is needed to improve thrombectomy modeling.

In this work, we aim to analyze the sensitivity of friction force between stent-vessel, stent-clot, and clot-vessel interaction. In addition, we have varied the retrieval speed to understand its impact on thrombectomy outcomes, on one hand as this can be varied by the operator during mechanical thrombectomies, and on the other hand as it corresponds to an important parameter for VMT simulations. The motivation of this study is to understand the sensitive range of friction coefficients between successful and unsuccessful thrombectomy outcomes and showcase the importance of estimating and validating accurate friction coefficients for MT. For this purpose, we have used two different realistic cerebral circulation models to reproduce the procedure with and without the presence of the clot. In both cases, the SR designs were replicated digitally. When the clot was present, its mechanical properties were calibrated based on experimental tests. We have also validated the numerical stent-vessel friction coefficients by comparing them with in vitro stent retrieval experiments.

## 2. Materials and methods

### 2.1. In vitro SR experiments

A benchtop circulatory flow loop, similar to Poulos et al. (2024), was developed to investigate SR behavior during MT retrieval (without clot) with different stent-vessel frictions [20]. This in vitro model consisted of a fluid reservoir, peristaltic pump, and a silicone ICA/MCA model (United Biologics), as shown in Fig. 1. To model SR MT, a Solitaire X (Medtronic) SR 6 × 40 mm within a Phenom 21 (Medtronic) microcatheter was passed through a hemostatic valve and navigated through the arterial model to the distal segment of the MCA, where the SR was slowly unsheathed from the microcatheter and deployed. The pushwire of the SR was then attached to a syringe pump (kdScientific), which functioned as a stepper motor to pull the SR at a constant speed of 2 mm/s. The test concluded when the SR was retracted through the entire arterial model and pulled through tortuous locations of the MCA and ICA. Videos of SR retrieval were recorded of each test to observe the SR's stretching and compressing behavior.

In vitro SR-MTs were performed with both water and a soap-water solution in the flow loop to vary the friction between the SR and the arterial model. To confirm that the soap-water solution created a more lubricated surface, the dynamic friction coefficients were compared between an SR and lubricated (with soap) silicone, and an SR and non-lubricated silicone. Experiments conducted by Elkhayyat et al. (2024) previously determined the coefficient of dynamic friction between a nitinol SR and silicone wetted with water (i.e. non-lubricated silicone) [19]. This friction testing protocol was repeated with a lubricated silicone surface using a custom testing apparatus developed for a Single Column Materials Testing System (Instron). A nitinol SR was then unfolded over the surface and pulled across the lubricated surface at 2 mm/s while the resultant pull forces were measured by the Instron's load cell. In comparison to the SR-non-lubricated dynamic friction coefficient of 0.73, the SR-lubricated dynamic friction coefficient was found to be 0.5.

### 2.2. In silico replica of the SR experimental tests

In the present work, for initial validation of stent-vessel behavior in our VMT simulations (without clot), we used a 3D model reconstructed from CT images of a silicone ICA/MCA experimental model used for the in-vitro SR-MT experiments described in 2.1 (Fig. 1). This model was discretized with 0.3 mm triangular rigid elements in ANSA Preprocessor v24.0 (BETA CAE System, Switzerland). The Solitaire X SR 6 × 40 mm was virtually created using MATLAB (R2020, MathWorks, USA) and discretized using 0.2 mm Hughes-Liu beam elements. The Solitaire stent has an open-end design and the stent material was modeled using shape memory alloy and material properties acquired from an experimental uniaxial tensile test [15]. The microcatheter, modelled as a 0.8 mm diameter tube, was discretized in ANSA with 0.15 mm triangular elements and it was assigned a linear elastic material. The experimental tests have been mimicked using the in silico model without blood clots, similar to the procedure shown in Fig. 2.

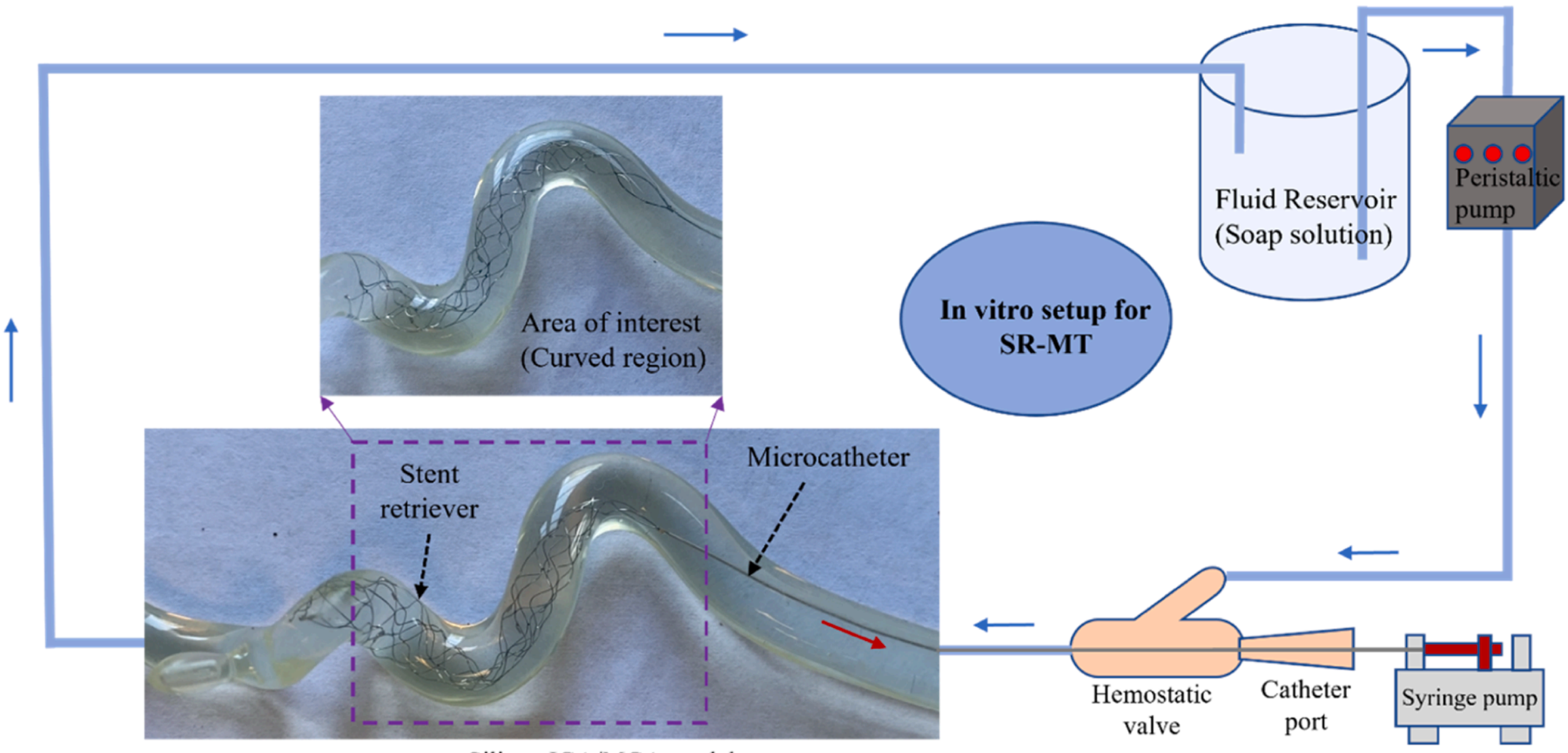

**Fig. 1.** Schematic of experimental SR-MT without clot to investigate SR behavior in curved regions of the model.

Mass proportional damping was adopted to improve stability. A selective mass scaling was adopted to fix the time-step size at7 $\times 10^{-7}$ s. The simulations were run with the finite-element commercial solver LS-DYNA R14 (ANSYS, Canonsburg, PA, USA) on 28 CPUs of an Intel Xeon64 with 250 GB RAM.

Quantitative in vitro and in silico comparison of stent compression has been performed by comparing stent compression at various time instances during retrieval. The stent compression is represented by percentage of stent compression and is defined by compressed SR diameter normalized with vessel diameter at curved region.

### 2.3. In silico MT

To model the MT, an idealized geometry of an average patient was used and discretized with 0.3 mm triangular elements. A 13.5 mm RBC-rich clot was placed in the distal M1, discretized with 0.2 mm tetrahedral elements, and modelled with a quasi-hyperelastic compressible foam formulation with a failure criterion based on a critical maximum principal stress value [21]. The same SR model used in the stent-only simulation was used for all the simulations. The simulation consists of four steps as shown in Fig. 2(b–e): (i) stent crimping inside the microcatheter, (ii) stent tracking to reach the occlusion site, (iii) stent deployment by pulling the microcatheter, and (iv) retrieval. More details about these steps can be found in our previous studies [15,17,21].

All contacts (stent-microcatheter, microcatheter-vessel, and stent-vessel) were defined as soft with a penalty. Soft penalty contact allows interaction between contacting surfaces and applies a penalty force to push them apart. Upon interpenetration of two surfaces, a restoring force proportional to the penetration depth is applied to resist further intrusion and enforce contact separation. As before, mass proportional damping was adopted to improve stability. In addition, a selective mass scaling was adopted to fix the time-step size at 7 $\times 10^{-7}$ s. After mesh independence analysis, the final mesh has been obtained and used for all the simulations. More details on the mesh independence analysis procedure can be found in our previously published article [15]. The simulations were run with the finite-element commercial solver LS-DYNA R14 (ANSYS, Canonsburg, PA, USA) on 28 CPUs of an Intel Xeon64 with 250 GB RAM.

A total of nine VMT simulations were performed with varying friction coefficients, as shown in Table 1. The friction coefficients between clot-vessel and clot-stent were taken from the in vitro study by Elkhayyat et al. with reported standard deviations used to determine the range for each [19]. Applied friction coefficients for stent-vessel were determined from our own VMT experiments to mimic in vitro data (shown in Fig. 3).

## 3. Results

### 3.1. Validation with in vitro MT for stent-vessel friction

Stent-vessel friction force is one of the important aspects of thrombectomy treatment and its outcome. Clinically, it has been observed that curved anatomy plays a decisive role in thrombectomy, and acutely curved vessels show negative outcomes of MT. In vitro experiments show the stretching and compression of stents in curved and tortuous anatomical regions. Using the in silico model, we have compared the stent's behavior in the cerebral vessel without a clot to understand the stent mechanics during retrieval. It has been observed that the stent gets stretched and compressed in the curved regions of the vessel, which could potentially impact the thrombectomy outcomes. Fig. 3 compares the stent's in vitro and in silico behaviors during retrieval. Fig. 3(a) compares the quantitative behavior of the stent with water as the in vitro fluid medium (top) and similar stent behavior is achieved by varying the stent-vessel friction using numerical simulations (bottom), as shown by red arrows. Stent compression between in vitro and in silico models shows good agreement, with a maximum percentage error of less than 10 %. The percentage compression in the in silico model varies from 21–25 % due to highly dynamic stent behavior in curved region. The coefficient of dynamic friction observed from the in vitro study was 0.73, while numerically this value needed to be between 0.3–0.4 to achieve similar stent behavior. The difference in friction coefficients between in silico and in vitro models may arise due to unknown factors during experiments or the absence of a fluid medium and rigid vessel walls during the in silico study. Also, contact force overestimation in the in silico model may show the difference in friction coefficient values between the in silico and in vitro models. The numerical code modeled contact as a soft penalty and a perfectly smooth surface by ignoring the surface irregularities at the micro-scale. While in vitro model has surface irregularities at the micro scale. This can be one of the causes of discrepancy in friction coefficients between in silico and in vitro models.

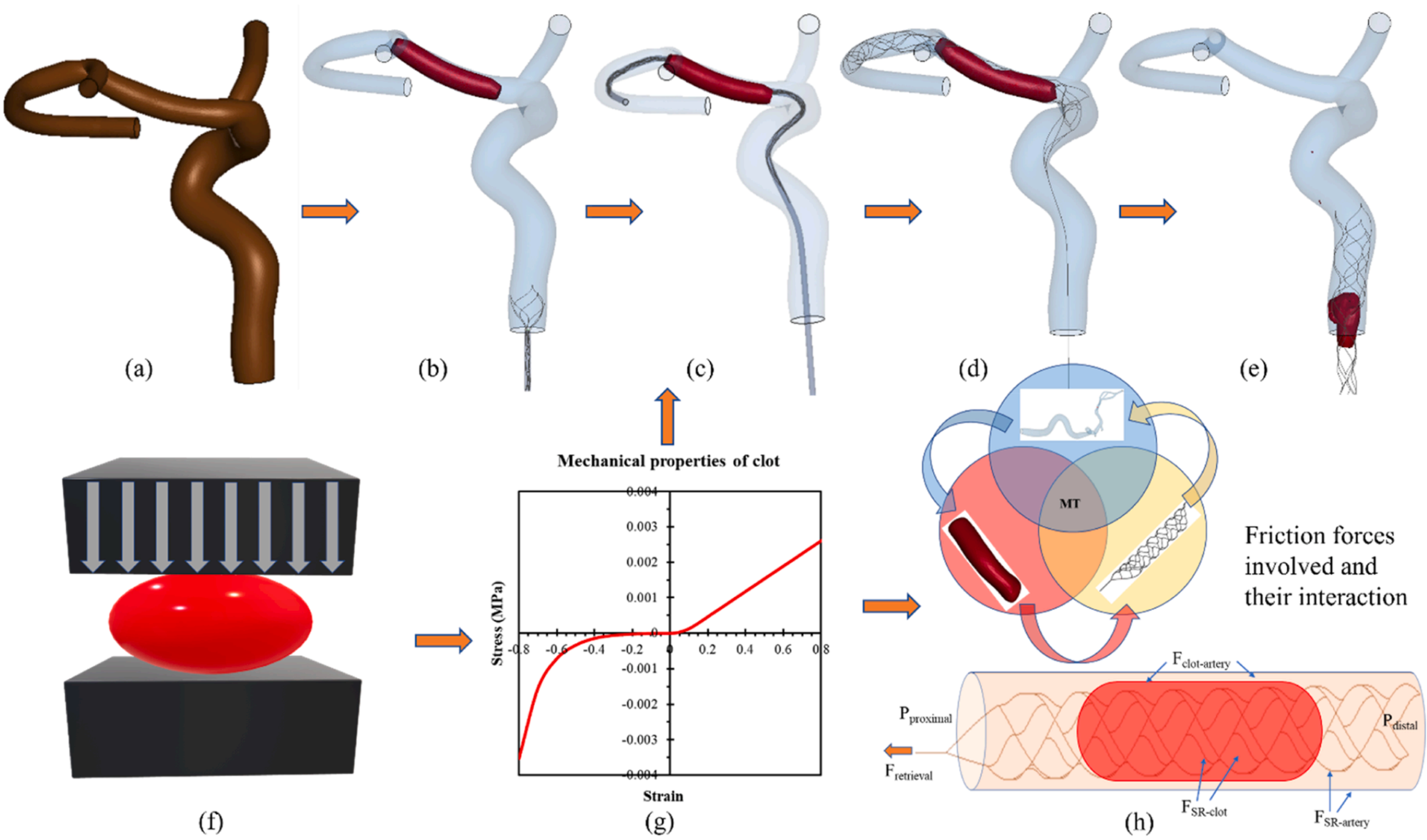

**Fig. 2.** Schematic of the pipeline used in the VMT simulations; (a) CAD model, (b) Stent crimping, (c) Stent tracking, (d) Stent deployment, (e) Clot retrieval, and (f) Compression test for the RBC-rich clot. (g) The stress-strain curve for the clot analogs from venous whole ovine blood was adapted by interpolating known curves [22]. (h) Diagram of friction forces involved in MT.

**Table 1**
Friction coefficients (RBC-rich clot) used for VMT simulations.

| Case | Retrieval Speed (mm/s) | Clot-vessel friction coefficient | | Clot-stent friction coefficient | | Stent-vessel friction coefficient | |
|---|---|---|---|---|---|---|---|
| | | Static | Dynamic | Static | Dynamic | Static | Dynamic |
| 1 | 20 | 0.46 | 0.23 | 0.10 | 0.03 | 0.10 | 0.10 |
| 2 | **2** | 0.46 | 0.23 | 0.10 | 0.03 | 0.10 | 0.10 |
| 3 | **350** | 0.46 | 0.23 | 0.10 | 0.03 | 0.10 | 0.10 |
| 4 | 20 | **0.35** | **0.09** | 0.10 | 0.03 | 0.10 | 0.10 |
| 5 | 20 | **0.57** | **0.37** | 0.10 | 0.03 | 0.10 | 0.10 |
| 6 | 20 | 0.46 | 0.23 | **0.09** | **0.02** | 0.10 | 0.10 |
| 7 | 20 | 0.46 | 0.23 | **0.11** | **0.04** | 0.10 | 0.10 |
| 8 | 20 | 0.46 | 0.23 | 0.10 | 0.03 | **0.15** | **0.15** |
| 9 | 20 | 0.46 | 0.23 | 0.10 | 0.03 | **0.20** | **0.20** |

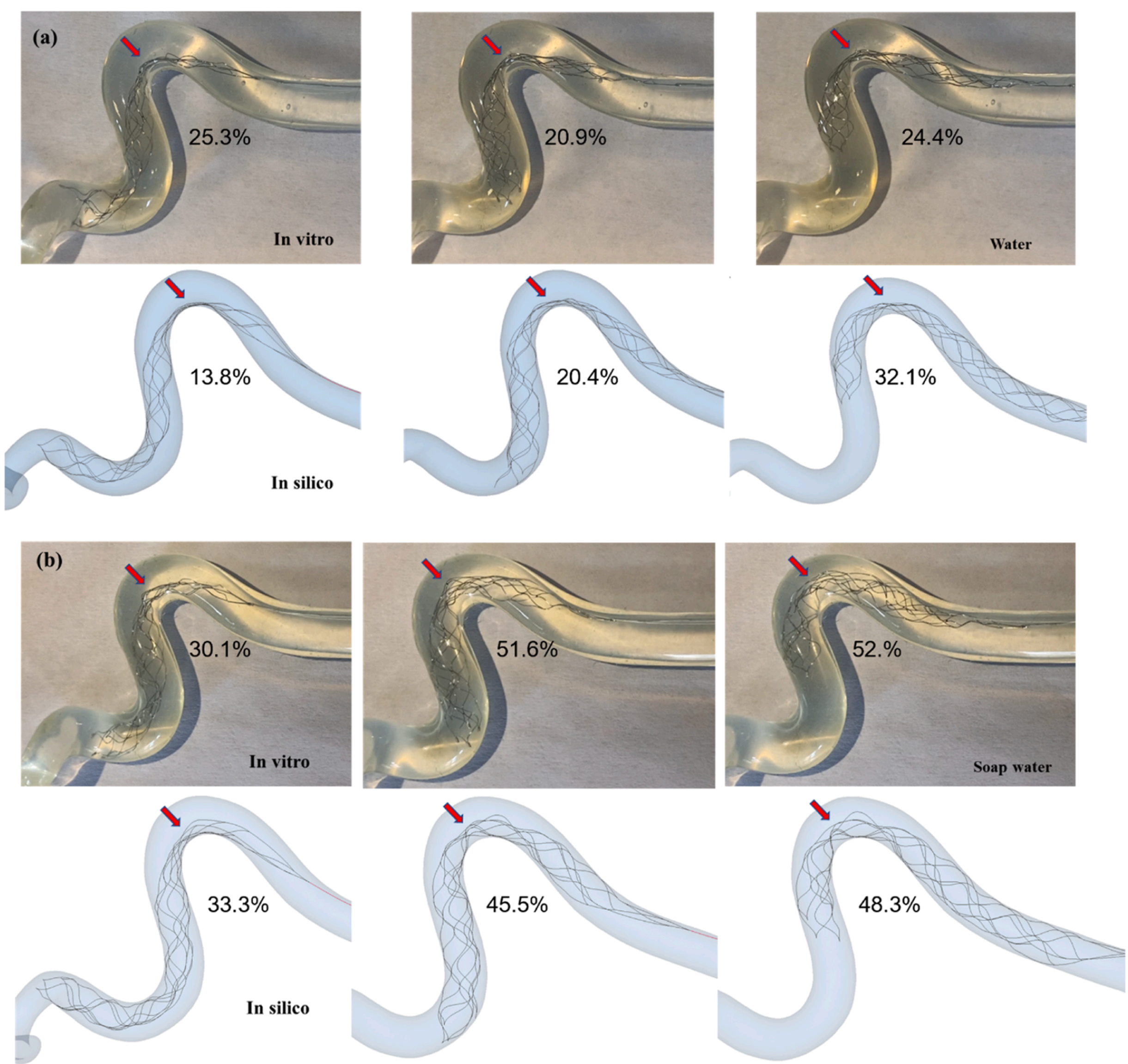

**Fig. 3.** Quantitative comparison of stent behavior during stent retrieval in MT using in vitro experiments with a silicone ICA/MCA model (United Biologics) and validating the stent behavior for in silico thrombectomy trials. The in vitro experiments were performed with both (a) water (in vitro friction coefficient of 0.73, numerical friction coefficient of 0.3) and (b) water-soap solution (in vitro friction coefficient of 0.5, numerical friction coefficient of 0.2).

Similarly, Fig. 3(b) shows the quantitative comparison of SR behaviors with the water-soap solution as the in vitro study fluid medium (top) and the in silico study (bottom). In vitro studies clearly show a significant reduction in dynamic friction coefficient from 0.73 to 0.5 with the addition of a few drops of soap into the water. To achieve similar stent behavior for the in silico experiment, the friction coefficient needed to be between 0.1–0.2. The percentage of stent compression is increased due to reduction in friction coefficient by soap solution instead of water. The quantitative comparison showed a good agreement between in vitro and in silico models. The maximum error between in silico and in vitro is about 10 %. To the best of authors knowledge, this is the first time we have tried to validate stent behavior qualitatively and quantitatively even after having highly dynamic scenario. We observed a consistent difference in friction coefficients between in vitro and in silico experiments. The use of a soap solution is a common practice in experimental models to get the SR behavior closer to the in vivo scenario [23,24]. Numerically, to replicate such behavior, however, the stent-vessel friction should be less than 0.2.

In vitro, the retrieval speed used was 2 mm/s, while in silico, the value was 20 mm/s. We performed 2 mm/s retrieval simulations previously but found them to be computationally expensive and not feasible for the number of total simulations required in this study. The stent compression for the retrieval speeds of 20 mm/s and 2 mm/s were found to be similar and therefore, we have compared the stent behavior to match the stent compression in the in vitro and in silico models using a retrieval speed of 20 mm/s. Please refer to supplementary material for comparison of SR compression at various retrieval speed and friction coefficient of 0.1.

### 3.2. *Effect of SR retrieval speed*

Fig. 4 shows the impact of variation in SR retrieval speed on the fragmentation of red clots and thrombectomy outcomes. In clinical practice, the retrieval mechanism is completely manual, and clinicians don't have accurate control over retrieval speed, but they can differentiate between 2 mm/s and 20 mm/s with the training and experience. Therefore, it is important to know the impact of variation in retrieval speed on blood clot behavior and thrombectomy outcomes. We have chosen the retrieval speed of 2 and 20 mm/s, which is very close to clinical settings, and an unrealistic 350 mm/s because it has been attempted in the past in silico simulations to reduce the computational time. By varying the retrieval speed, the friction coefficients for clot-stent (FS = 0.1, FD = 0.03), clot-vessel (FS = 0.46, FD = 0.23), and stent-vessel (FS = FD = 0.2) were kept constant for all three cases. At a lower retrieval speed (2 mm/s), the clot gets completely entrapped in the stent and retrieved successfully (Fig. 4a). However, a slight increase in retrieval speed changes the thrombectomy outcome to partial retrieval, as shown in Fig. 4(b). At a lower retrieval speed, the clot gets entrapped within the middle part of the stent retriever. While higher retrieval speed fragments the clot from the beginning of the retrieval and leads to delayed entrapment, as shown in Fig. 4(c).

The results clearly show that higher retrieval speed contributes to RBC-rich clot fragmentation and thrombectomy failure. The fragmentation of the clot at a higher retrieval speed may be a result of stent compression at curved regions. Due to the stretching of stent, the wire-like structure penetrates through clot and leads to fragmentation. At higher retrieval speed, the clot entraps at the distal part of the SR, leading to dislodgment and failure in the thrombectomy outcome. Higher retrieval speed may also injure the endothelial cells in cerebral vasculature and lead to secondary complications. The impact of higher retrieval speed on vessel injury is unknown and needs to be explored further. The results also show that simulating VMT at unrealistic retrieval speeds (to reduce computing time) may overestimate clot fragmentation for RBC-rich clots (see Fig. 4(c)), and early dislodgment of white clots (shown in supplementary information Fig. 1).

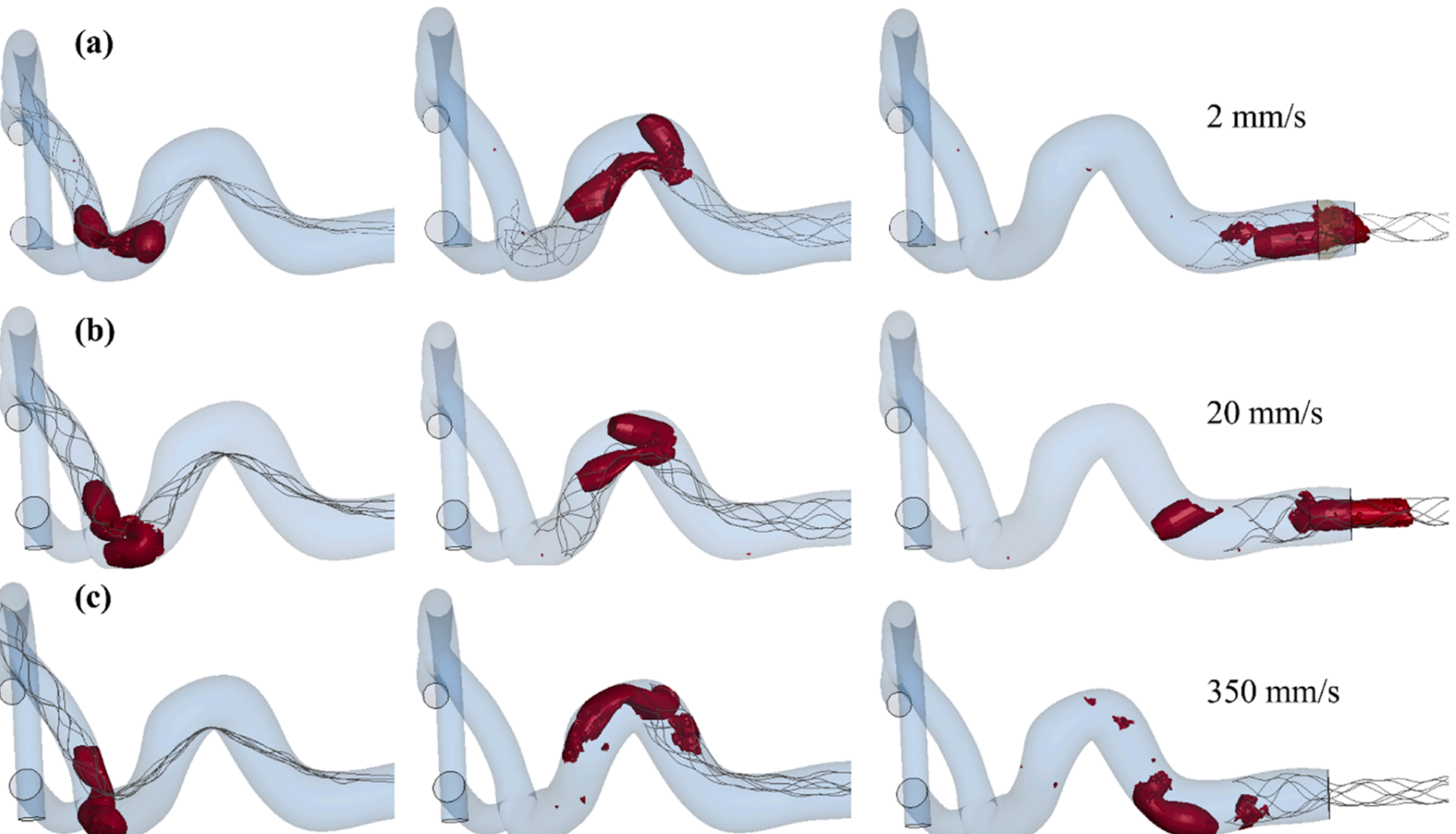

**Fig. 4.** Variation of retrieval speed during VMT with speeds of (a) 2 mm/s, (b) 20 mm/s, and (c) 350 mm/s. Friction coefficients for Clot-stent (FS = 0.1, FD = 0.03), clot-vessel (FS = 0.46, FD = 0.23), and stent-vessel (FS=FD=0.2) were kept constant for all cases.

### 3.3. Effect of clot-vessel friction

We have performed VMT simulations with varying clot-vessel friction coefficients within the standard deviations observed from in vitro studies. By varying the clot-vessel friction, the friction coefficients for clot-stent (FS = 0.1, FD = 0.03) and stent-vessel (FS=FD=0.2) were kept constant for all three cases. Similarly, the retrieval speed of 20 mm/s is used for all three cases. The use of a lower retrieval speed (2 mm/s) is computationally expensive while the highest retrieval speed (350 mm/s) leads to extreme stent compression and clot fragmentation, as discussed in Section 3.2. Fig. 5 shows the effect of clot-vessel friction on the retrieval mechanism. It can be seen that with an increase in friction coefficient, clot retrieval becomes more difficult. Clot retrieval was successful in the minimum friction coefficient case, and the thrombectomy outcome was positive. In this case, the other friction forces between stent-clot and stent-vessel are almost similar to clot-vessel friction. It might be possible that a similar range in friction forces leads to perfect retrieval, as shown in Fig. 5(a). A slight increase in the clot-vessel friction shows minor fragmentation of the clot from the SR at the proximal curvature, which leads to successful retrieval with minor fragmentation, as shown in Fig. 5(b). Further, Fig. 5(c) shows an increase in the clot-vessel friction coefficient to the maximum value reported in Table 1. It shows the breaking of a clot in the proximal artery region, at the post-acutely curved region. The retrieval in the higher friction case was partial and a large part of the clot dislodged from the stent near to the proximal curved vessel region.

The clot fragmentation was observed in higher clot-vessel friction cases, while no fragmentation was observed in lower friction coefficient cases. The fragmentation of the clot with higher friction could be caused by the bending of the clot at the proximal curvature, as shown in Fig. 5 (c). Bending of the clot was observed due to the interplay between higher clot-vessel friction and momentum of the stent due to its pulling force. The higher friction coefficient between the clot surface and vessel induces maximum principal stress in the clot above the rupture limit leading to the breaking of the clot into small fragments. However, these results are only for RBC-rich clots, and fibrin-rich clots might show different behaviors in such a case.

### 3.4. Effect of clot-stent friction

Fig. 6 shows the impact of friction coefficient variations between the clot and stent. By varying the clot-stent friction, the stent-vessel (FS = FD = 0.1), and clot-vessel (FS = 0.46, FD = 0.23) friction coefficients were kept constant for all three cases. The retrieval speed of 20 mm/s is used for all the simulations. The lower value of clot-stent friction indicates an entrapment of the clot in the stent, but slippage is observed due to lower friction. In the lower friction case, the thrombectomy outcome shows an unsuccessful retrieval, as shown in Fig. 6(a). Further, an increase in the friction force between the clot and the stent shows an improved entrapment of the clot within the stent, which leads to the successful pulling of the clot, as shown in Fig. 6(b). Fig. 6(c) shows the results for the highest friction coefficient case and it causes clot fragmentation during the early retrieval stage. Due to higher friction between the clot and the stent, clot fragmentation is observed at the first curvature in the distal region. The smaller fragmented clot is difficult to retain in the stent cage and slips off midway of the vessel length during the retrieval phase. In investigating clot-stent interactions, very low friction causes clot slippage while higher friction leads to clot fragmentation.

### 3.5. Effect of stent-vessel friction

Fig. 7 shows the effect of varying friction coefficients for stent-vessel interaction on thrombectomy outcome. The clot-stent (FS = 0.1, FD = 0.03) and clot-vessel (FS = 0.46, FD = 0.23) friction coefficients were kept constant for all three cases. The retrieval speed is kept constant at 20 mm/s for all three simulations. In the lower friction coefficient case, the clot gets entrapped in the stent, but fragmentation is observed near the proximal region, as shown in Fig. 7(a). A slight increase in the friction coefficient causes clot fragmentation at the distal region of the vessel as well. Due to an increase in the friction coefficient to 0.15, minor compression of stent occurs at the curved region. The stent penetrates through the clot and leads to increased fragmentation, as shown in Fig. 7(b). Further, an increase in friction causes an increase in stent compression at curved vessel regions. Due to compression of the stent, the proximal end of the clot disengages from the stent and the clot is twisted at the distal curvature, as shown in Fig. 7(c).

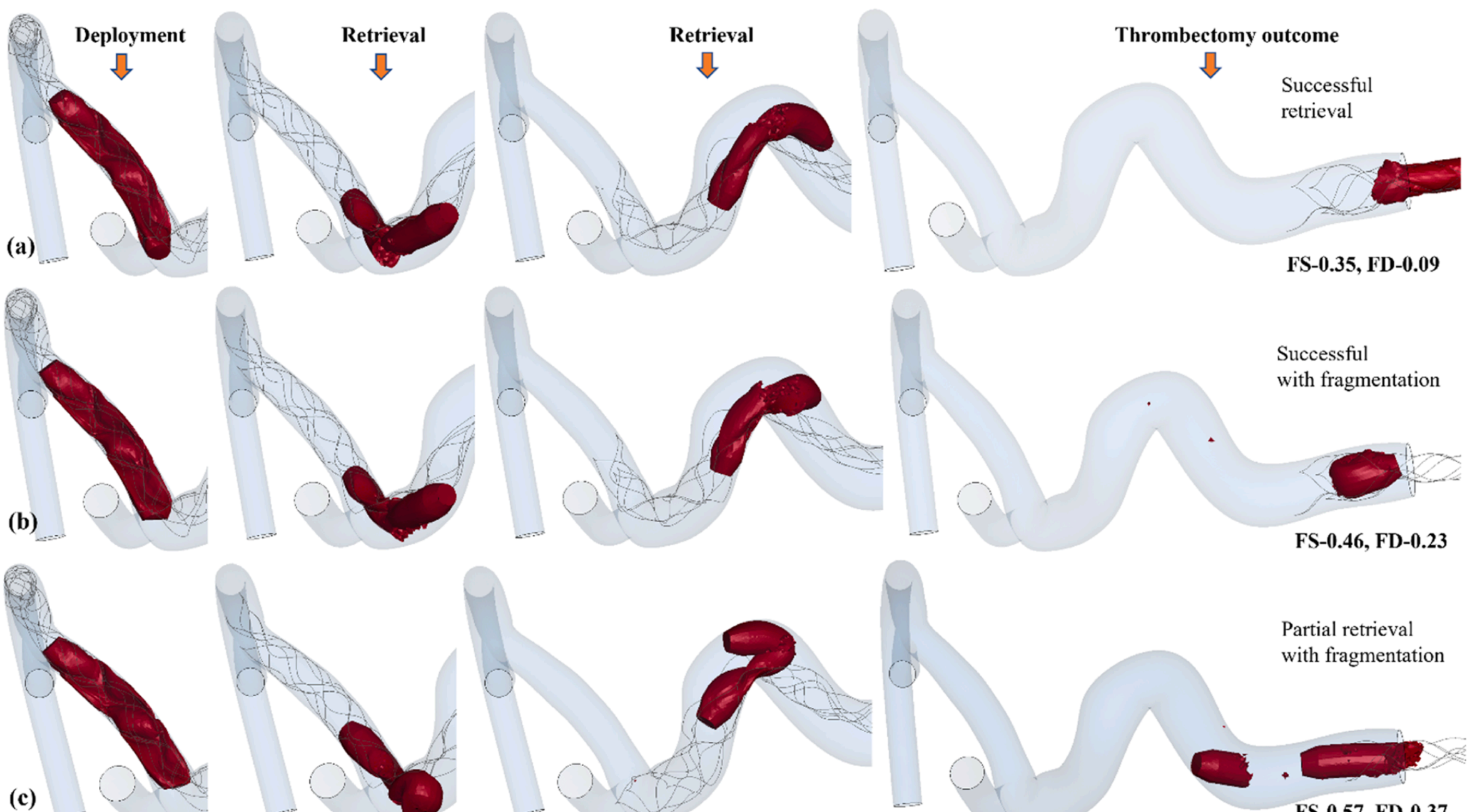

**Fig. 5.** Variation of static (FS) and dynamic (FD) friction coefficients for clot-vessel interaction within the range of minimum and maximum values observed from in vitro studies; (a) Minimum friction, (b) Mean friction, and (c) Maximum friction. Stent-vessel (FS = FD = 0.1) and Stent-clot (FS = 0.1, FD = 0.03) friction coefficients were kept constant for all three cases.

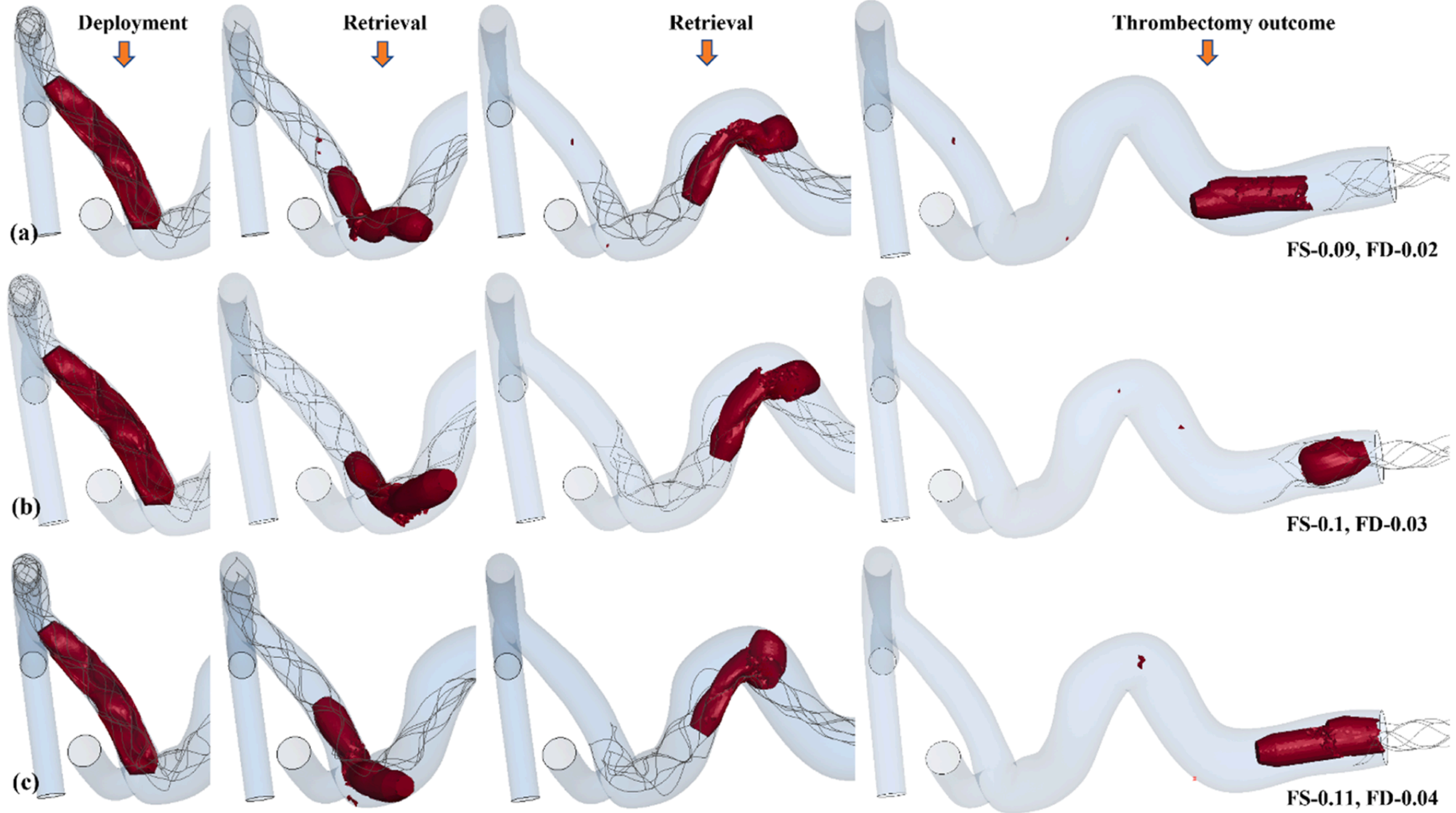

**Fig. 6.** Variation of static (FS) and dynamic (FD) friction coefficients for clot-stent interaction within the range of minimum and maximum values observed from in vitro studies; (a) Minimum friction, (b) Mean friction, and (c) Maximum friction. Stent-vessel (FS = FD = 0.1) and clot-vessel (FS = 0.46, FD = 0.23) friction coefficients were kept constant for all three cases.

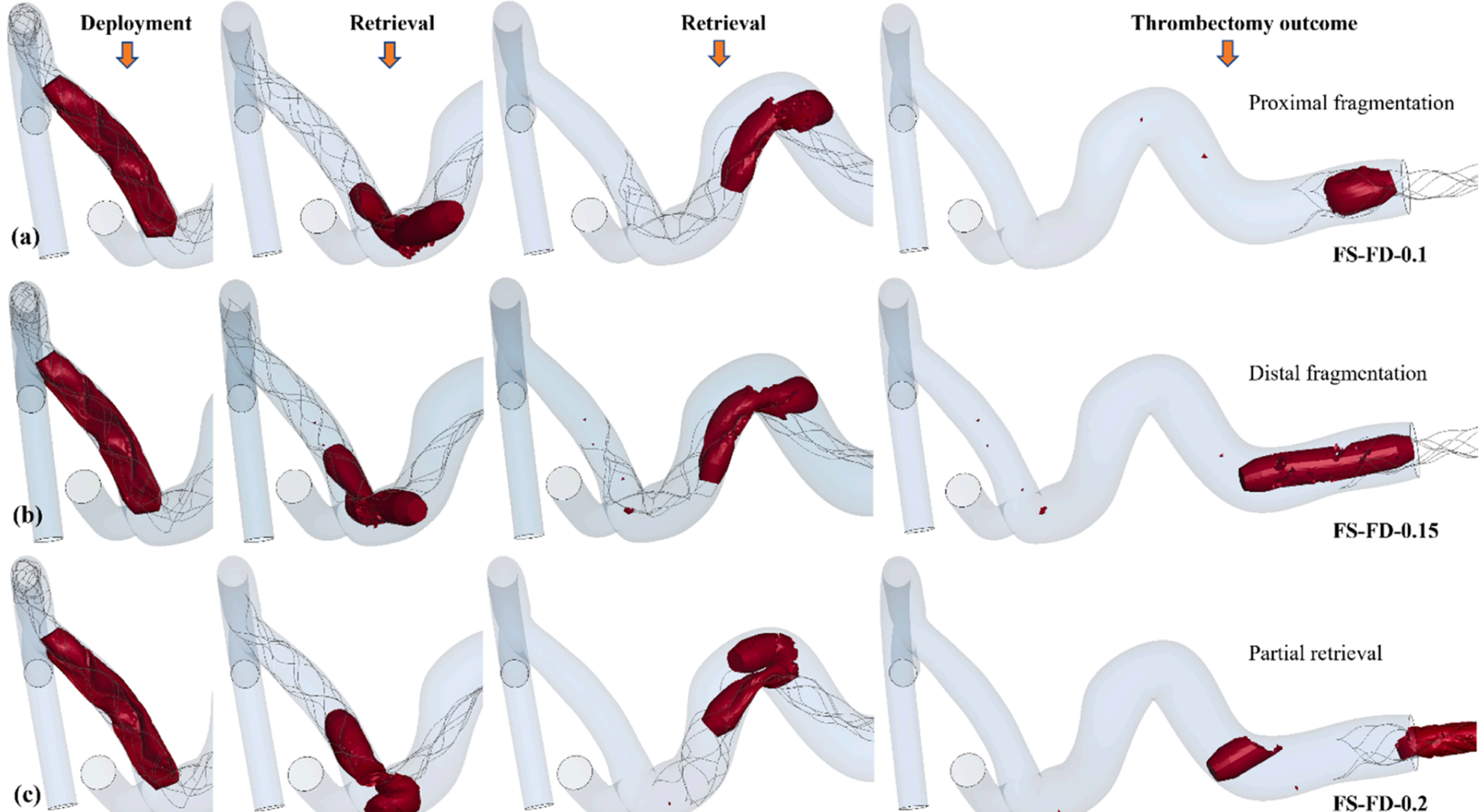

**Fig. 7.** Variation of static (FS) and dynamic (FD) friction coefficients for stent-vessel interaction within the range of minimum and maximum values observed from in vitro studies; (a) Minimum friction, (b) Mean friction, and (c) Maximum friction. Clot-stent (FS = 0.1, FD = 0.03) and clot-vessel (FS = 0.46, FD = 0.23) friction coefficients were kept constant for all three cases.

The twisting of the clot is observed at the proximal curvature as well and as the clot slides against the vessel wall. Due to interplay between stent momentum and frictional forces between the clot and vessel wall leads to breaking of the clot into two parts. In the higher stent-vessel friction case, clot fragmentation is observed and thrombectomy outcome leads to partial retrieval, as seen in Fig. 7(c). We showed previously (Fig. 3) that VMT stent-vessel friction coefficients to mimic the in vitro experiments need to lie between 0.1–0.2. Generally, it is well known that stents become compressed in curved regions of the vasculature during surgical treatment, but no knowledge of the degree of compression is available.

## 4. Discussion

Multiple studies have shown the effectiveness of MT and its success in recanalization rates [5,6]. However, most cases faced higher failure rates or multiple attempts to remove the embolus using a SR [4]. The failure in complete retrieval of the clot leads to a secondary embolism in the distal arteries of cerebral circulation. Multiple SR attempts injure the inner lumen of the artery, and it can generate secondary embolus due to wall remodeling. Later, fibrin-rich clots were found more difficult to retrieve compared to RBC-rich clots [25,26]. These clinical observations open new doors for in vitro analysis of clot properties in MT.

Various studies attempted in vitro MT analysis and found it beneficial in understanding the failure mechanism [12,13]. An important study by Gunning et al. found that the failure in retrieval is related to the friction force between the clot and the vessel [18]. It has been observed that fibrin-rich clots have significantly higher friction coefficients compared to RBC-rich clots. Due to higher friction with fibrin-rich clots, they show higher failure rates during retrieval. Gunning et al. calculated the static friction coefficients between the clot-artery, but static friction only plays a crucial role until the clot is dislodged. Once the clot gets entrapped in SR, the dynamic friction plays a dominant role during retrieval. A recent article by Elkhayyat et al. proposed a way of calculating static and dynamic friction coefficients for clot-artery and stent-vessel [19]. For the first time, they calculated dynamic friction coefficients for clot-vessel, stent-vessel, and SR-vessel using in vitro models. The experimental calculation of friction forces involves complex procedures and shows large standard deviations in the results during the repeatability of experiments. In vitro models restrict us from parametric sensitivity analysis of friction forces and its influence on the outcome of thrombectomy success. However, in silico models provide flexibility for friction force sensitivity analysis.

The motivation of the present study is to understand the sensitivity of friction force for clot-vessel, clot-stent, and stent-vessel individually on thrombectomy outcomes. We have adapted static and dynamic friction coefficient ranges from the in vitro analysis by Elkhayyat et al. (2024) and implemented those in our in silico model [19]. The clot-vessel friction variation significantly impacts thrombectomy outcomes, as it is a decisive force during detachment or dislodging of the clot during retrieval. Higher clot-vessel friction shows complete entrapment during deployment and dislodging of the clot from the stent by fragmentation during retrieval. During retrieval, the clot slipped off from the SR due to dominant clot-vessel friction in comparison with a stent-clot friction force. While at lower clot-vessel friction the clot remains entrapped in the stent cage and is retrieved successfully. Each individual's clot might be different and may show different friction behavior based on its composition. It has been accepted that clot can be a combination of fibrin-rich and RBC-rich behaviors. To the best of the author's knowledge, no available work has analyzed large data for retrieved clots and their friction behavior in each case. Such a study will guide our understanding of friction forces and their implications for thrombectomy success by performing large data set patient-specific simulations. In silico simulations will gain more confidence after performing it on large data and comparing those with surgical outcomes. The inclusion of uncertainty quantification of friction coefficients into the in silico models enhances its credibility and acceptance in clinical practice.

The clot-stent friction has been calculated during in vitro experiments and is considered secondary in terms of its impact on thrombectomy outcomes. It was observed that clot-stent friction does not vary significantly for RBC-rich and fibrin-rich clots. These observations showed its ignorance towards thrombectomy success, but the current in silico model varied the thrombectomy outcome by a slight variation in clot-stent friction. The variation in clot-stent friction was within the range of standard deviation observed in prior in vitro studies. The current work concluded that very low and very high friction between the stent and clot is not favorable for thrombectomy success. It has been attempted that increasing the clot-stent friction is possible using surface modification to the existing commercial SRs [27]. Such surface modifications will be beneficial for improving the clot entrapment and improve MT outcomes by reducing distal embolization. Lower values of friction lead to slippage of clots, while higher values lead to fragmentation. The microscopic images of the entrapped clot and stent might give a deeper understanding of the slippage and fragmentation of RBC-rich clots [28], but such studies on wider data sets are not available and need to be addressed.

It has been observed clinically that failure in thrombectomy often occurs in acutely curved vessels [29,30]. Past research has been dedicated to understanding the effect of vessel tortuosity on MT using in vitro models. Poulos et al. (2024) observed that SR removal forces increased with an increase in the vessel tortuosity [20]. Further, Tsuto et al. (2024) found that narrow and acutely curved vessels have higher frictional forces [31]. The stent-vessel interaction is one of the major areas of research for in vitro models. In vitro models observed stent stretching in acutely curved vessels, which reduces the cross-sectional area of the vessel covered by the stent cage [32,33]. Similar stretching behavior of stents in curved vessels was observed through in vivo imaging [34]. Due to the stretching of the stent, the clot loses contact with the stent cage and gets dislodged in curved regions. The stent-vessel friction force has been ignored in previous in silico studies, but it shows a significant dominance in thrombectomy outcomes. For the first time, we observed this behavior with in silico models and it agrees well with prior in vitro and in vivo studies. With the increase in stent-vessel dynamic friction, the chances of clot dislodgment from the stent are higher. We found that higher friction leads to dislodgment of fragmented clots during retrieval in the proximal region of the cerebral vessel. This is related to stent compression in curved regions, which reduces the possibility of clot retention in the stent.

The effect of retrieval speed on MT has been explored using in vitro and in vivo studies [35,36]. They found that fast retrieval (20–30 mm/s) increased the success ratio of thrombectomy for RBC-rich and fibrin-rich clots. However, they did not investigate the post-thrombectomy effect of fast retrieval on vessel injury. For the first time, we have explored the effect of retrieval speed on the fragmentation of clots and thrombectomy outcomes using in silico modeling. The use of a lower retrieval speed (2 mm/s) is close to the clinical settings, but it is highly expensive in terms of computational time. While the highest retrieval speed (350 mm/s) shows unrealistic behavior of stent in curved regions and may lead to incorrect estimates of the procedure outcome. We recommend setting the numerical retrieval speed close to the values found in the clinic. To reduce the computational time we recommend not to exceed the maximum retrieval speed of 20 mm/s, as it gives better thrombectomy prediction. Fast retrieval (20–30 mm/s) might be helpful for the removal of fibrin-rich clots, but a decision has to be taken based on the severity of injury caused due to fast retrieval speed. Further in vitro and in silico studies are needed to better understand the success ratio of thrombectomy on fibrin-rich clots using fast retrieval.

## 5. Limitations

This work has major limitations, such as rigid vessel walls and the absence of blood flow from collateral circulation. Further work is needed to perform virtual mechanical thrombectomy using flexible vessels. Friction force analysis has been performed with RBC-rich clots and fibrin-rich clots. The fibrin-rich clots show a negative outcome in all the cases due to its even higher friction coefficients and are included in the supplementary material. Further research is needed to explore fibrin-rich clot retrieval using advanced stent retriever design or techniques. The use of flexible artery walls might change the friction force and needs to be addressed in future studies. Also, collateral blood flow from the distal region might help dislodge the clot more easily, which requires a separate study and is out of the scope of the current work. Despite these limitations, in silico models play a crucial role in understanding hidden mechanics and can contribute to improvements in thrombectomy treatments and medical device developments. The findings from this work can be beneficial for stent retriever manufacturers to optimize MT procedure. The knowledge of higher retrieval speed affect negatively on thrombectomy outcomes could be helpful to physicians to improve the thrombectomy outcomes.

## 6. Conclusions

In silico models are valuable tools for the prediction of intervention techniques and testing the performance of medical devices. The US Food and Drug Administration recommends the use of in silico models by following their guidelines [37,38]. In this view, our past studies have performed an applicability analysis to establish the credibility of thrombectomy simulations [15,16]. The current friction coefficient sensitivity analysis can be useful in the future for performing in vitro and in silico analysis on large datasets of patients. Results show that uncertainty in the friction coefficient is a determinant factor in modeling MT. Varying the friction coefficient, even within a limited range may lead to different thrombectomy outcomes such as successful, partial retrieval, and unsuccessful attempts. Hence, accurate modeling of thrombectomy should consider uncertainty quantification of this effect when reporting the potential outcome of an intervention. Further work has to be performed on calculating in vivo friction coefficients or getting more confidence in VMT by performing patient-specific cases on large patient data sets. More simulations need to be performed on large datasets in close collaboration with clinicians to gain the credibility of in silico thrombectomy simulations. For the first time in the literature, this work investigated the importance of accurate modeling of friction coefficients as it is one of the deciding factors in thrombectomy outcomes. The higher retrieval speed affects the thrombectomy outcome negatively and it is advised to use physiological retrieval speed range in the thrombectomy simulations.

## Funding

This project has received funding from the European Union's Horizon 2020 research and innovation program under the Marie Sklodowska-Curie grant agreement No. 101,104,493.

FM, GL, and JFRM are partially supported by the European Union's Horizon Europe research and innovation program, grant number 101,136,438.

## Ethics statement

This study did not require ethical approval as it involved the use of publicly available datasets that do not contain identifiable information about individuals. Informed consent was therefore not applicable. The research adhered to all applicable laws and regulations regarding data use.

## CRediT authorship contribution statement

**Mahesh S. Nagargoje:** Writing – original draft, Visualization, Validation, Project administration, Methodology, Investigation, Funding acquisition, Formal analysis, Data curation, Conceptualization. **Virginia Fregona:** Visualization, Validation, Software, Resources, Methodology, Formal analysis, Data curation. **Giulia Luraghi:** Writing – review & editing, Supervision, Resources, Project administration, Methodology, Investigation, Funding acquisition, Formal analysis, Conceptualization. **Francesco Migliavacca:** Writing – review & editing, Supervision, Resources, Project administration, Methodology, Investigation, Funding acquisition, Formal analysis, Conceptualization. **Demitria A Poulos:** Visualization, Validation, Methodology, Data curation. **Bryan C Good:** Writing – review & editing, Visualization, Validation, Supervision, Methodology, Investigation, Formal analysis. **Jose Felix Rodriguez Matas:** Writing – review & editing, Supervision, Resources, Project administration, Methodology, Investigation, Funding acquisition, Formal analysis, Conceptualization.

## Declaration of competing interest

The authors declare that they have no known competing financial interests or personal relationships that could have appeared to influence the work reported in this paper.

## Acknowledgments

None.

## Supplementary materials

Supplementary material associated with this article can be found, in the online version, at doi:10.1016/j.cmpb.2025.109018.